\documentstyle[12pt]{article}

\def\o{\over}
\def\b{\begin{equation}}
\def\e{\end{equation}}
\def\l{\label}
\def\kpnn{$K^+\rightarrow\pi^+\nu\bar\nu$\ }
\def\kpn{K^+\rightarrow\pi^+\nu\bar\nu}
\def\klpnn{$K_L\rightarrow\pi^0\nu\bar\nu$\ }
\def\klpn{K_L\rightarrow\pi^0\nu\bar\nu}

\def\imlt{{\rm Im}\lambda_t}
\def\relt{{\rm Re}\lambda_t}
\def\relc{{\rm Re}\lambda_c}

\def\r#1{(\ref{#1})}

\begin{document}
\thispagestyle{empty}
\begin{flushright}
 MPI-PhT/94-19 \\
 TUM-T31-59/94 \\
 April 1994
\end{flushright}
\vskip1truecm
\centerline{\Large\bf $\sin 2\beta$ from $K\to\pi\nu\bar\nu$
   \footnote[1]{\noindent
   Supported by the German
   Bundesministerium f\"ur Forschung und Technologie under contract
   06 TM 732 and by the CEC science project SC1--CT91--0729.}}
\vskip1truecm
\centerline{\sc Gerhard Buchalla {\rm and} Andrzej J. Buras}
\bigskip
\centerline{\sl Max-Planck-Institut f\"ur Physik}
\centerline{\sl  -- Werner-Heisenberg-Institut --}
\centerline{\sl F\"ohringer Ring 6, D-80805 M\"unchen, Germany}
\vskip0.6truecm
\centerline{\sl Technische Universit\"at M\"unchen, Physik Department}
\centerline{\sl D-85748 Garching, Germany}

\vskip1truecm
\centerline{\bf Abstract}
We point out that the measurement of just the two branching fractions
\linebreak
$B(\kpn)$ and $B(\klpn)$ can in a theoretically clean manner
determine
$\sin 2\beta$
almost independently of $m_t$ and $V_{cb}$. This allows to
obtain an interesting relation between the CP asymmetry
$A_{CP}(\psi K_S)$ in B physics and the branching ratios for these
two rare K decays. The recently calculated next-to-leading order
QCD corrections improve the accuracy of this analysis.
We find typically $\Delta\sin 2\beta=\pm 0.11$ provided
$B(\kpn)$ and $B(\klpn)$ are measured within $\pm 10\%$ accuracy.
With decreasing uncertainty in $\Lambda_{\overline{MS}}$ and $m_c$
this error could be reduced to
$\Delta\sin 2\beta < \pm 0.10$.
The determination of $\sin 2\alpha$ and
$\sin 2\gamma$ on the other hand is rather poor. However
respectable determinations of the Wolfenstein parameter
$\eta$ and of $\mid V_{td}\mid$ can be obtained.
\vfill
\newpage

\pagenumbering{arabic}

The search for the unitarity triangle is one of the important
targets of contemporary particle physics. It is well known that
CP asymmetries in $B^0\to f$, where $f$ is a CP eigenstate will play
an important role in this enterprise.
The CP asymmetry in the decay $B^0_d\rightarrow \psi K_S$ allows
 in the standard model
a direct measurement of the angle $\beta$ in the unitarity triangle
without any theoretical uncertainties
\cite{NI}. Similarly the decay
$B^0_d\rightarrow \pi^+ \pi^-$ gives the angle $\alpha$, although
 in this case strategies involving
other channels are necessary in order to remove hadronic
uncertainties related to penguin contributions
\cite{GL1,GL2,NQ1,NQ2,ADKL}.
The determination of the angle $\gamma$ from CP asymmetries in neutral
B-decays is more difficult but not impossible \cite{AKL,FL}.
\\
On the other hand it has been pointed out in \cite{BH} that
measurements of $B(\kpn)$ and $B(\klpn)$ could also determine the
unitarity triangle completely provided $m_t$ and $V_{cb}$ are known.
It should be stressed that these two decays are theoretically very
clean. \klpnn is dominated by short distance loop diagrams involving
the top quark and proceeds almost entirely through direct CP violation
\cite{LI}. \kpnn is CP conserving and receives contributions from
both internal top and charm exchanges.
The recent calculation
of next-to-leading QCD corrections to these decays \cite{BB1,BB2,BB3}
consi\-derably reduced the theoretical uncertainty
due to the choice of the renormalization scales present in the
leading order expressions \cite{DDG}. Since the relevant hadronic matrix
elements of the weak current $\bar {s}\gamma_{\mu}(1- \gamma_{5})d$
can be measured in the leading
decay $K^+ \rightarrow \pi^0 e^+ \nu$, the resulting theoretical
expressions for $B(\kpn)$ and $B(\klpn)$ are
only functions of the CKM parameters, the QCD scale
 $\Lambda \overline{_{MS}}$
 and the
quark masses $m_t$ and $m_c$.
The long distance contributions to
$K^+ \rightarrow \pi^+ \nu \bar{\nu}$ have been
considered in \cite{RS,HL,LW} and found to be very small: two to three
orders of magnitude smaller than the short distance contribution
at the level of the branching ratio.
The long distance contributions to \klpnn are negligible as well.

In view of this theoretical development and anticipating improvements
in the knowledge of $V_{cb}$ and $m_t$ we would like to make a
closer look on this determination of the unitarity triangle.
In particular we want to point out that $\sin 2\beta$ can be entirely
expressed in terms of $B(\kpn)$ and $B(\klpn)$ with only
a very weak dependence
on $V_{cb}$ and $m_t$. Consequently \kpnn and \klpnn offer an
independent clean
determination of $\sin 2\beta$ which can be confronted with
the one possible in $B^0\to\psi K_S$. Combining these two ways of
determining $\sin 2\beta$ we obtain an interesting relation
between $A_{CP}(\psi K_S)$, $B(\kpn)$ and $B(\klpn)$ which must be
satisfied in the standard model. Any deviation from this relation
would signal new physics.
\\
The present experimental upper bound on $B(\kpn)$ is
$5.2\cdot 10^{-9}$ \cite{AT}. An improvement by one order of
magnitude, closing the gap to standard model expectations,
is planned at AGS in Brookhaven for the coming years.
The present upper bound on $B(\klpn)$ from Fermilab experiment E731
is $2.2\cdot 10^{-4}$ \cite{GR}. FNAL-E799 expects to reach an accuracy
of ${\cal O}(10^{-8})$. Eventually a sensitivity of ${\cal O}(10^{-12})$
and ${\cal O}(10^{-11})$, the level necessary according to standard
model predictions,
could be achieved at KAMI \cite{AR} and
KEK \cite{ISS} respectively.
\\
Our discussion of the Cabibbo-Kobayashi-Maskawa matrix
will be based on the standard parametrization \cite{PDG},
which can equivalently be rewritten in terms of the Wolfenstein
parameters ($\lambda$, $A$, $\varrho$, $\eta$) through the
definitions
\b\l{wop}
s_{12}\equiv\lambda \qquad s_{23}\equiv A \lambda^2 \qquad
s_{13} e^{-i\delta}\equiv A \lambda^3 (\varrho-i \eta)      \e
Due to the resulting simplifications, the Wolfenstein
parametrization \cite{WO} is particularly useful when
an expansion in $\lambda=|V_{us}|=0.22$ is performed.
Including next-to-leading terms in $\lambda$ \cite{BLO}
implies that the point A in the reduced unitarity triangle of fig. 1
defined through
\b\l{ut}
\bar\varrho + i \bar\eta\equiv -{V_{ud}V^\ast_{ub}\o
 V_{cd}V^\ast_{cb}}   \e
is with an error of less than 0.1\%
given by $\bar\varrho=\varrho (1-\lambda^2/2)$ and
$\bar\eta=\eta (1-\lambda^2/2)$ and not by ($\varrho$, $\eta$) as
usually found in the literature. Working in the Wolfenstein
parametrization such a treatment is required if
we aim at a determination of the unitarity triangle with improved
precision.

The explicit expression for $B(\kpn)$ is given as follows
\b\l{bkpn}
B(\kpn)=\kappa\cdot\left[\left({\imlt\o\lambda^5}X(x_t)\right)^2+
\left({\relc\o\lambda}P_0(K^+)+{\relt\o\lambda^5}X(x_t)\right)^2
\right]            \e
\b\l{kap}
\kappa={3\alpha^2 B(K^+\to\pi^0e^+\nu)\o 2\pi^2\sin^4\Theta_W}
 \lambda^8=4.64\cdot 10^{-11}   \e
Here $x_t=m^2_t/M^2_W$, $\lambda_i=V^\ast_{is}V_{id}$. $\lambda_c$
is real to a high accuracy. The function $X$ can be written as
\b\l{xex0}
X(x) = \eta_X \cdot \frac{x}{8} \left[ - \frac{2+x}{1-x} + \frac{3x
-6}{(1-x)^2} \ln x \right] \qquad \quad \eta_X = 0.985  \e
where $\eta_X$ is the NLO correction calculated in \cite{BB2}.
With $m_t\equiv\bar m_t(m_t)$ the QCD factor $\eta_X$ is practically
independent of $m_t$. Next
\b\l{p0k}
P_0(K^+)=\frac{1}{\lambda^4}\left[\frac{2}{3} X^e_{NL}+\frac{1}{3}
 X^\tau_{NL}\right]   \e
with $X^l_{NL}$ calculated in \cite{BB3}. We remark that in writing
$B(\kpn)$ in the form of \r{bkpn} a negligibly small term
$\sim (X^e_{NL}-X^\tau_{NL})^2$ has been omitted.
Numerical values of $P_0(K^+)$
are given in table 1 where $m_c\equiv\bar m_c(m_c)$. Note that our
definition of $P_0(K^+)$ differs by a factor $-(1-\lambda^2/2)$ from
that used in \cite{BH}.
\begin{table}
\begin{center}
\begin{tabular}{|c|c|c|c|}\hline
&\multicolumn{3}{c|}{$P_0(K^+)$}\\ \hline
$\Lambda_{\overline{MS}}$, $m_c$ [$GeV$]
&1.25&1.30&1.35\\ \hline
0.20&0.409&0.443&0.479\\ \hline
0.25&0.393&0.427&0.462\\ \hline
0.30&0.376&0.410&0.445\\ \hline
0.35&0.359&0.393&0.428\\ \hline
\end{tabular}
\end{center}
\centerline{}
{\bf Table 1:} The function $P_0(K^+)$ for various
$\Lambda_{\overline{MS}}$ and $m_c$.
\end{table}

Furthermore
\b\l{bklpn}
B(K_L\to\pi^0\nu\bar\nu)=\kappa_L\cdot\left({\imlt\o\lambda^5}X(x_t)
  \right)^2    \e
\b\l{kapl}
\kappa_L=\kappa{\tau(K_L)\o\tau(K^+)}=1.94\cdot 10^{-10}  \e
It is evident from \r{bkpn} and \r{bklpn} that, given $B(\kpn)$
and $B(\klpn)$, one can extract both $\imlt$ and $\relt$. We find
\b\l{imre}
\imlt=\lambda^5{\sqrt{B_2}\o X(x_t)}\qquad
\relt=-\lambda^5{{\relc\o\lambda}P_0(K^+)+\sqrt{B_1-B_2}\o X(x_t)} \e
where we have introduced the "reduced" branching ratios
\b\l{b1b2}
B_1={B(\kpn)\o 4.64\cdot 10^{-11}}\qquad
B_2={B(\klpn)\o 1.94\cdot 10^{-10}}   \e
Using the standard parametrization of the CKM matrix and \r{ut} we
then obtain $\bar\varrho$, $\bar\eta$ in terms of $\relt$ and $\imlt$
\b\l{rho}
\bar\varrho={\sqrt{1+4 s_{12}c_{12}\relt /s^2_{23}-
(2 s_{12}c_{12}\imlt /s^2_{23})^2}-
1+2 s^2_{12}\o 2 c^2_{23} s^2_{12}}   \e
\b\l{eta}
\bar\eta={c_{12}\imlt\o s_{12}c^2_{23}s^2_{23}}   \e
Up to the excellent approximations that $V_{cd}V^\ast_{cb}$ is real
(error below 0.1\%) and
$c_{13}=1$ (error less than $10^{-5}$)
\r{rho}, \r{eta} are {\it exact\/} relations. Together with \r{imre}
they determine the unitarity triangle of fig. 1 in terms of
$B(\kpn)$ and $B(\klpn)$.
\\
Once $\bar\varrho$ and $\bar\eta$ have been determined this way we can
calculate several quantities of interest as functions of $B_1$ and
$B_2$ \r{b1b2}. In particular
\b\l{sin}
r_s=r_s(B_1, B_2)={1-\bar\varrho\o\bar\eta}=\cot\beta \qquad
\sin 2\beta=\frac{2 r_s}{1+r^2_s}   \e
\b\l{vubtd}
|V_{td}|=A\lambda^3 \sqrt{(1-\bar\varrho)^2 +\bar\eta^2}   \e
and $\sin 2\alpha$, $\sin 2\gamma$ for which formulae are given
in \cite{BLO}.
In order to see the implications of \r{rho}, \r{eta} more clearly,
it is useful to expand these equations in powers of the Wolfenstein
parameter $\lambda$. $\relt$ and $\imlt$ are of the order
${\cal O}(\lambda^5)$. Therefore we define
\b\l{ab}
a=-{\relt\o\lambda^5}\qquad b={\imlt\o\lambda^5}   \e
and obtain (recall \r{wop})
\b\l{rhol}
\bar\varrho=1-{a\o A^2}(1-{\lambda^2\o 2})-\lambda^2
 {a^2+b^2\o A^4} + {\cal O}(\lambda^4)   \e
\b\l{etal}
\bar\eta={b\o A^2}(1-{\lambda^2\o 2}) + {\cal O}(\lambda^4)   \e
Since
\b\l{rsl}
r_s= \cot\beta={1-\bar\varrho\o\bar\eta}={a\o b}+
 \lambda^2{a^2+b^2\o b A^2} + {\cal O}(\lambda^4)   \e
we observe using \r{imre} and \r{ab} that to leading order in $\lambda$
the angle $\beta$ is completely {\it independent\/} of $m_t$ and
$V_{cb}$ (or $A$). The dependence on these parameters enters only at
order ${\cal O}(\lambda^2)$. Due to this suppression the
determination of $\sin 2\beta$ from $B(\kpn)$ and $B(\klpn)$ will
be rather insensitive to the values of $m_t$ and $V_{cb}$.
\\
On the other hand
the time integrated CP violating asymmetry
in $B^0_d\to\psi K_S$ is given by
\b\l{acp}
A_{CP}(\psi K_S)=-\sin 2\beta {x_d\o 1+x^2_d}  \e
where $x_d=\Delta m/\Gamma$ gives the size of $B^0_d-\bar B^0_d$
mixing. Combining \r{sin} and \r{acp} we obtain an
interesting connection between rare K decays and B physics
\b\l{kbcon}
{2 r_s(B_1,B_2)\o 1+r^2_s(B_1,B_2)}=-A_{CP}(\psi K_S){1+x^2_d\o x_d}   \e
which must be satisfied in the standard model. We stress that except
for $P_0(K^+)$ given in table 1 all quantities in \r{kbcon} can
be directly measured in experiment and that this relationship is
almost independent of $m_t$ and $V_{cb}$.

We would like to compare the formulae given here with the discussion
of \kpnn presented in \cite{BLO}. In this paper approximate
expressions have been proposed for $\relt$ and $\imlt$ as functions
of the Wolfenstein parameters. These approximations are equivalent
to using \r{etal} (without ${\cal O}(\lambda^4)$-terms) for $\bar\eta$
and to replace \r{rhol} for $\bar\varrho$ by
\b\l{rholb}
\bar\varrho=1-{a\o A^2}(1-{\lambda^2\o 2})^{-1}   \e
While the approximation for $\bar\eta$ is very precise
(error below 0.1\%), $\bar\varrho$ given in \r{rholb} may
deviate by typically $\Delta\bar\varrho=0.02$
from the exact expression in \r{rho}. Comparing \r{rholb} with
\r{rhol} we observe that \r{rholb} neglects some  ${\cal O}(\lambda^2)$
terms, which offers the possibility
to obtain very simple, and still rather accurate,
explicit relations between $\bar\varrho$,
$\bar\eta$ and the $K\to\pi\nu\bar\nu$ branching ratios. In fact,
using \r{imre}, \r{ab}, \r{etal}, \r{rholb}, we may write
($\sigma=(1-\lambda^2/2)^{-2}$)
\b\l{rhetb}
\bar\varrho=1+{P_0(K^+)-\sqrt{\sigma(B_1-B_2)}\o A^2 X(x_t)}\qquad
\bar\eta={\sqrt{B_2}\o\sqrt{\sigma} A^2 X(x_t)}  \e
and
\b\l{cbb} \cot\beta=r_s(B_1, B_2)=
\sqrt{\sigma}{\sqrt{\sigma(B_1-B_2)}-P_0(K^+)\o\sqrt{B_2}} \e
Obviously within this approximation the small dependence of $\beta$
on $m_t$ and $V_{cb}$ drops out.

\centerline{}
We now turn to a brief numerical investigation of the
phenomenology discussed above. In \r{kap} and \r{kapl} we have used
\cite{PDG}
\b\l{las}
\lambda=0.22\quad\alpha=1/128\quad \sin^2\Theta_W=0.23 \e
\b\l{btklp}
B(K^+\to\pi^0e^+\nu)=4.82\cdot 10^{-2}\qquad
\tau(K_L)/\tau(K^+)=4.18   \e
As an illustrative example, let us consider the following scenario.
We assume that the branching ratios are known to within $\pm 10\%$
\b\l{bkpkl}
B(\kpn)=(1.0\pm 0.1)\cdot 10^{-10}\qquad
B(\klpn)=(2.5\pm 0.25)\cdot 10^{-11}   \e
Next we take ($m_i\equiv\bar m_i(m_i)$)
\b\l{mtcv}
m_t=(170\pm 5)GeV\quad m_c=(1.30\pm 0.05)GeV\quad
V_{cb}=0.040\pm 0.001  \e
where the quoted errors are quite reasonable if one keeps in mind
that it will take at least ten years to achieve the accuracy
assumed in \r{bkpkl}. The value of $m_t$ in \r{mtcv} is in the ball
park of the most recent results of the CDF collaboration \cite{CDF}.
Finally we use
\b\l{lamuc}
\Lambda_{\overline{MS}}=(0.20-0.35)GeV\qquad
\mu_c=(1-3)GeV       \e
where $\mu_c$ is the renormalization scale present in the
analysis of the charm contribution. Its variation gives an
indication of the theoretical uncertainty involved in the calculation
\cite{BB3}.
In comparison to this error we neglect the effect of varying
$\mu_W={\cal O}(M_W)$, the high energy matching scale at which
the W boson is integrated out, as well as the very small scale
dependence of the top quark contribution.
As reference parameters we use the central values in \r{bkpkl} and
\r{mtcv} and $\Lambda_{\overline{MS}}=0.3GeV$, $\mu_c=m_c$.
The results that would be obtained in such a scenario for $\sin 2\beta$,
$\bar\eta$, $|V_{td}|$ and $\bar\varrho$ are collected in table 2.
\begin{table}
\begin{center}
\begin{tabular}{|c||c||c|c|c|c||c|}\hline
&&$\Delta(BR)$&$\Delta(m_t, V_{cb})$&$\Delta(m_c,\Lambda_{\overline{MS}})
$&$\Delta(\mu_c)$&$\Delta_{total}$\\ \hline
$\sin 2\beta$&$0.60$&$\pm 0.06$&$\pm 0.00$&$\pm 0.03$&$\pm 0.02$&
$\pm 0.11$\\ \hline
$\bar\eta$&$0.33$&$\pm 0.02$&$\pm 0.03$&$\pm 0.00$&$\pm 0.00$&
$\pm 0.05$\\ \hline
$|V_{td}|/10^{-3}$&$9.3$&$\pm 0.6$&$\pm 0.6$&$\pm 0.5$&$\pm 0.4$&
$\pm 2.1$\\ \hline
$\bar\varrho$&$0.00$&$\pm 0.08$&$\pm 0.09$&$\pm 0.06$&$\pm 0.04$&
$\pm 0.27$\\ \hline
\end{tabular}
\end{center}
\centerline{}
{\bf Table 2:} $\sin 2\beta$, $\bar\eta$, $|V_{td}|$ and
$\bar\varrho$ determined
from \kpnn and \klpnn for the scenario described in the text together
with the uncertainties related to various parameters.
\end{table}
There we have also displayed separately the associated, symmetrized
errors
($\Delta$) coming from the uncertainties in the branching ratios,
$m_t$ and $V_{cb}$, $m_c$ and $\Lambda_{\overline{MS}}$, $\mu_c$,
as well as the total uncertainty.
We briefly mention a few interesting points concerning the
determination of $\sin 2\beta$. The accuracy to which $\sin 2\beta$
can be obtained from $K\to\pi\nu\bar\nu$ is, in our example,
comparable to the one expected in determining $\sin 2\beta$ from
CP asymmetries in B decays \cite{BAB,AL} prior to LHC experiments
\cite{CA}.
The largest partial uncertainty comes
from the branching ratios themselves ($\pm 0.06$). Due to the very
weak dependence of $\sin 2\beta$ on $m_t$ and $V_{cb}$ the error
from these input parameters is negligibly small ($<0.003$).
Note that the theoretical uncertainty represented by $\Delta(\mu_c)$,
which ultimately limits the accuracy of the analysis, is quite
moderate. This reflects the clean nature of the $K\to\pi\nu\bar\nu$
decays. However the small uncertainty of $\pm 0.02$ is only achieved by
including next-to-leading order QCD corrections \cite{BB2,BB3}.
In the leading logarithmic approximation
the corresponding error would amount to
$\pm 0.05$, larger than the one coming from $m_c$ and
$\Lambda_{\overline{MS}}$.
We expect that in the coming years the uncertainty in $m_c$ and
$\Lambda_{\overline{MS}}$ could be reduced so that
$\Delta\sin 2\beta < \pm 0.10$ is quite conceivable.
\\
Similar comments apply to $\bar\eta$ and $|V_{td}|$. Unfortunately
$\sin 2\alpha$ and $\sin 2\gamma$ can not be determined from
$K\to\pi\nu\bar\nu$ very reliably unless the input parameters are
known with even higher precision than considered in our example.
The errors on $\sin 2\alpha$ and $\sin 2\gamma$ are generally larger
than found in \cite{BLO}. This is also reflected in the sizable
range for $\bar\varrho$ shown in table 2.
\\
We have checked that a similar picture emerges for different
branching ratios with the same relative uncertainties assumed.

Let us finally summarize the main aspects of this letter.
\\
It is well known that the rare kaon decays \kpnn and \klpnn are
theoretically very well under control and are therefore expected
to play a potentially important role in tests of standard model
flavordynamics \cite{LV,WW,RW}.
In particular these two decays alone can, at least
in principle, fix the unitarity triangle completely \cite{BH}.
Furthermore the expressions for the branching ratios are now available
at next-to-leading order in QCD \cite{BB2,BB3}. For these reasons we found it
interesting to look into the perspective for obtaining parameters
of the unitarity triangle from $K\to\pi\nu\bar\nu$. The most
promising quantities are $\bar\eta$, $|V_{td}|$ and $\sin 2\beta$.
We stress the following points:
\begin{itemize}
\item The measurement of only \kpnn and \klpnn with an error of
$\pm 10\%$ can yield $\sin 2\beta$ with an uncertainty
comparable to that of the determination from CP asymmetries in
B decays prior to LHC experiments.
We are aware, however, of the fact that these measurements, in
particular of $\klpn$, are very challenging experimentally.
\item $\sin 2\beta$ determined from $K\to\pi\nu\bar\nu$ decays is
only very weakly dependent on $m_t$ and $V_{cb}$. The uncertainties
in these parameters are therefore greatly reduced and a
clean relation between $B(\kpn)$, $B(\klpn)$, $A_{CP}(\psi K_S)$
and $x_d$ can be established.
\item Moreover, the measurement of \klpnn would provide a
uniquely clear evidence for
{\it direct\/} CP violation.
\end{itemize}
In this letter we have emphasized the physics potential of the
rare decays \klpnn and $\kpn$. While we certainly share the great
enthusiasm in the physics community to measure CP asymmetries in
neutral B decays, we are convinced that in addition
all efforts should be made
to measure the branching ratios of these theoretically very clean
rare kaon decays.

\section*{Figure Caption}

{\bf Fig. 1:} Unitarity triangle in the complex
$(\bar\varrho, \bar\eta)$ plane.

\vfill\eject

\end{document}